\def\percmsq{\rm {cm$^{-2}$}}

\def\kms{\rm {km\,s$^{-1}$}}
\def\H2O{H$_2$O}
\def\18wat{H$_2^{18}$O}
\def\wat17{H$_2^{17}$O}
\def\amm{NH$_{3}$}

\documentclass{aa}
\usepackage{graphicx}
\usepackage{txfonts}
\usepackage{natbib}
\bibpunct{(}{)}{;}{a}{}{,}

\begin{document}

\title{Ground-state ammonia and water in absorption towards Sgr B2} 

\author{E.S.~Wirstr\"om\inst{1} \and P.~Bergman\inst{1,2} \and
J.H.~Black\inst{1} \and \AA.~Hjalmarson\inst{1} \and
B.~Larsson\inst{3} \and A.O.H.~Olofsson\inst{1,4} \and
P.J.~Encrenaz\inst{5} \and E.~Falgarone\inst{6} \and
U.~Frisk\inst{7} \and M.~Olberg\inst{1} \and Aa. Sandqvist\inst{3} }    

\offprints{P.~Bergman, \\ \email{per.bergman@chalmers.se}}

\institute{Onsala Space Observatory, Chalmers University of
Technology, SE - 43992 Onsala, Sweden \and European Southern
Observatory, Alonso de Cordova 3107, Vitacura, Casilla 19001,
Santiago, Chile \and Stockholm Observatory, AlbaNova, SE
- 10691 Stockholm, Sweden \and GEPI, Observatoire de Paris, CNRS, 5
Place Jules Janssen, 92195 Meudon, France \and LERMA \& UMR 8112
du CNRS, Observatoire de Paris, 61 Av. de l'Observatoire, 75014 Paris,
France \and LERMA \& UMR 8112 du CNRS, \'Ecole Normale Sup\'erieure,
24 rue Lhomond, 75005 Paris, France \and Swedish Space Corporation, PO
Box 4207, SE - 17104 Solna, Sweden}   

\date{Received  /  Accepted}

\abstract
{Observations of transitions to the ground-state of a
molecule are essential to obtain a complete picture of its
excitation and chemistry in the interstellar medium, especially in
diffuse and/or cold environments. For the important interstellar
molecules \H2O\ and \amm, these ground-state transitions are heavily
absorbed by the terrestrial atmosphere, hence not observable
from the ground.}
{We attempt to understand the chemistry of nitrogen, oxygen, and their
important molecular forms, \amm\ and \H2O\
in the interstellar medium of the Galaxy.}
{We have used the Odin\thanks{Odin is a Swedish-led satellite project funded
jointly by the Swedish National Space Board (SNSB), the Canadian Space
Agency (CSA), the National Technology Agency of Finland (Tekes), and
the centre National d'\'Etudes Spatiales (CNES, France). The Swedish
Space Corporation (SSC) was the industrial prime contractor and is
also responsible for the satellite operation.} submillimetre-wave
satellite telescope to observe the ground state transitions of
ortho-ammonia and ortho-water, including their $^{15}$N, $^{18}$O, and
$^{17}$O isotopologues, towards Sgr B2. The extensive simultaneous velocity
coverage of the observations, $>$500~\kms, ensures that we can probe
the conditions of both the warm, dense gas of the molecular cloud Sgr B2
near the Galactic centre, and the more diffuse gas in the
Galactic disk clouds along the line-of-sight.} 
{We present ground-state \amm\ absorption in seven distinct velocity
features along the line-of-sight towards Sgr B2. We find a nearly linear
correlation between the column densities of \amm\ and CS, and a
square-root relation to N$_2$H$^+$. The 
ammonia abundance in these diffuse Galactic disk clouds is estimated to be
about 0.5--1$\times10^{-8}$, similar to that observed for diffuse
clouds in the outer Galaxy. On the basis of the detection of \18wat\
absorption in the 3~kpc arm, and the absence of such a feature in the \wat17\ spectrum, we conclude that the water abundance is around
10$^{-7}$, compared to $\sim$10$^{-8}$ for \amm. 
The Sgr B2 molecular cloud itself is seen in absorption in \amm,
$^{15}$\amm, \H2O, \18wat, and \wat17, with emission superimposed on
the absorption in the main isotopologues. The non-LTE excitation
of \amm\ in the environment of Sgr B2 can be explained without
invoking an unusually hot (500 K) molecular layer. A hot
layer is similarly not required to explain the line profiles of the
$1_{1,0}$$\gets$$1_{0,1}$ transition from H$_2$O and its isotopologues.  
The relatively weak $^{15}$\amm\ absorption in the Sgr B2
molecular cloud indicates a high [$^{14}$N/$^{15}$N] isotopic ratio
$>600$. The abundance ratio of \18wat\ and \wat17\ is found to be
relatively low, 2.5--3. These results together indicate
that the dominant nucleosynthesis process in the Galactic centre is CNO
hydrogen burning.
}
{}

\keywords{Astrochemistry -- ISM: molecules -- ISM: abundances --
Submillimeter -- Galaxy: disk}

\maketitle


\section{Introduction}
Less than 100 pc from the dynamical centre of the Galaxy lies its most massive
molecular cloud complex, Sgr B2.  
Embedded in the gas along a north-south line are three radio
continuum peaks called north (N), middle/main (M) and south (S). These 
peaks are separated from each other by approximately 50\arcsec, which
corresponds to projected linear separations of 1.9 pc at a distance
of 8 kpc.  Sgr B2(S) is the
weakest of the three and (N) is the most intense source at longer
wavelengths, $\lambda \geq 800\;\mu$m. Sgr B2(M) dominates at shorter
wavelengths. The molecular lines of Sgr B2 are seen in both emission
and absorption against the continuum sources
\citep[e.g.,][]{Nummelin, Polehampton07}. The hot core of
Sgr B2(N) displays an extreme chemical richness exhibiting strong emission
lines of many complex carbon-bearing molecules. In addition, line-of-sight
clouds associated with molecular gas concentrations in the Galactic
plane produce absorption lines towards Sgr B2 in several
atomic and molecular species, e.g. \ion{H}{i} 
\citep{Garwood}, \ion{O}{i} and \ion{C}{ii} \citep{Vastel}, H$_2$CO
\citep{Zuckerman,Whiteoak,Wadiak}, CS \citep{Greaves}, H$_2$O
\citep{Neufeld}, CH \citep{Whiteoak85,PolehamptonCH}, and OH
\citep{Robinson,PolehamptonOH}. 

We present observations of ground state ortho-ammonia and 
ortho-water, including their $^{15}$N, $^{18}$O, and $^{17}$O
isotopologues, towards Sgr B2. The submm-wave spectra cover more
than 500~\kms\ in Doppler velocity and thus trace absorption in both the Sgr B2 molecular cloud complex and several line-of-sight 
Galactic clouds.


\section{Observations and data reduction} \label{obssect}
The submillimetre wave spectroscopy satellite Odin has observed
absorption in the ground-state rotational line
$(J,K)$\,=\,(1,0)\,$\gets$\,(0,0) of \amm\ at 572.498~GHz as well as the 
ground-state rotational transition
$J_{K_-,K_+}$\,=\,$1_{1,0}$\,$\gets$\,$1_{0,1}$ of \H2O at 556.936~GHz,
\18wat at 547.676~GHz, and \wat17 at 552.021~GHz towards Sgr~B2. The
satellite was pointed within 3\arcsec\ of $\alpha$~=~17$^{\rm
h}$47$^{\rm m}$19\fs7, $\delta$~=~$-$28\degr\,22\arcmin\,51\farcs4
(J2000), which corresponds to a position between Sgr B2(M) and (N) and 
encompasses both these cores within the projected beam of 2\arcmin, 
full width at half-maximum (FWHM). 
The pointing uncertainty was $\leq$15\arcsec, not large enough to
affect the signal strengths significantly. 

The \amm\ and \wat17\ observations were carried out during August
(\amm: 30 orbits, 10 hours on-source) and September (\wat17: 60
orbits, 18 hours on-source) 2003, while the \H2O\ observations
comprised of 23 orbits (7 hours on-source) during February 2005. All
these data were acquired in position-switching mode (PSW) where the
spacecraft was reoriented to an off-position at ($-$1800\arcsec,
$+$1750\arcsec), in cycles of two minutes, to measure
signal-free reference spectra. 

The \18wat\ observations were carried out in two separate campaigns:
during September 2003 (60 orbits, 16 hours on-source) and October 2006 
(100 orbits, 28 hours on-source).   
The 2003 data were acquired in sky-switching mode, meaning
that pairs of on-source/off-source spectra were measured in cycles of 20
seconds by means of side-looking, large-beam sky mirrors and a chopper
wheel. In 2006, PSW was employed as described above. 

In all observations, the chopper wheel method was used for calibration 
\citep{Frisk,Olberg}. The resulting system temperatures were
typically 3200--3700~K near the tuning centres, but sometimes varied
significantly across the band, being up to a factor of two higher in
short segments in peripheral parts of the passband.
The data were recorded with an acousto-optical spectrometer (AOS)
giving a $\sim$0.3~\kms\ velocity resolution over $>$540~\kms\
bandwidth at all four frequency settings. In the case of the PSW data, the
effective velocity resolution is somewhat lower ($\sim$0.4~\kms) due
to the uncompensated change of satellite line-of-sight
velocity during the on-integrations. 

An additional 25 orbits of \18wat\ data were recorded using another
spectrometer in 2002 and -- while not included here due to stability issues
-- show good general agreement with the data acquired later.


\section{Results} \label{Res}
\begin{figure}
\resizebox{\hsize}{!}{\includegraphics{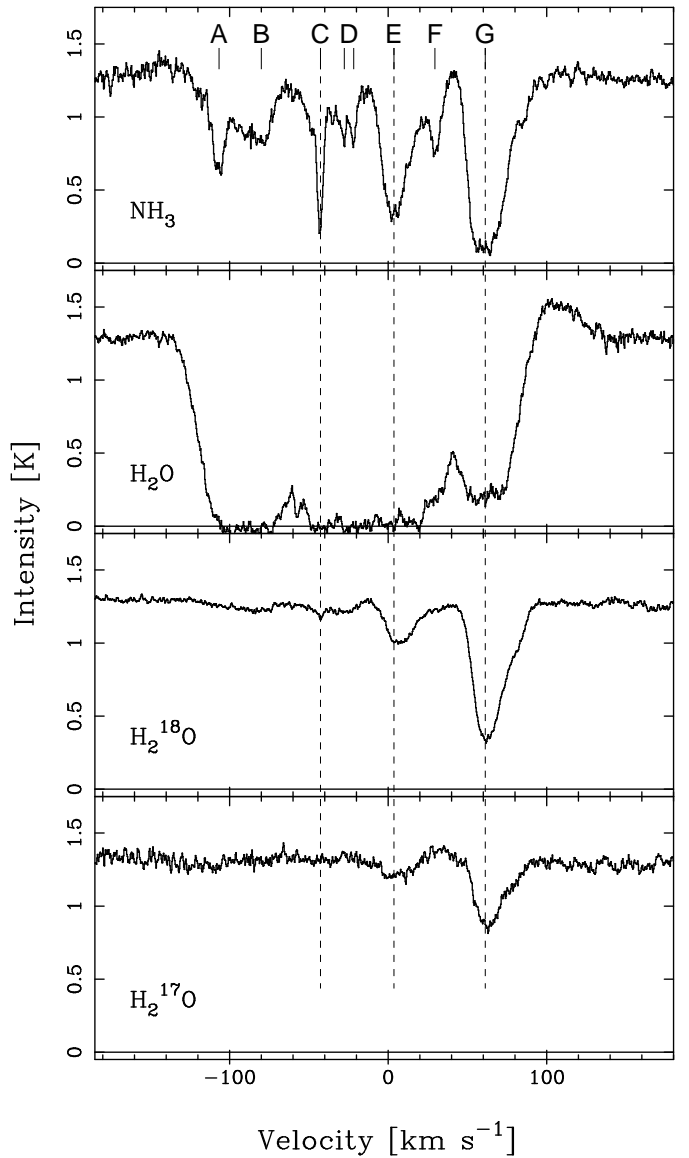}}
\caption{Part of the Odin spectra towards Sgr B2. The absorption
features marked {\bf{A}}--{\bf{G}} are identified in
Table~\ref{feattab}. The radial velocity is indicated with respect
to the local standard of rest and to the rest frequencies listed in 
Sect.~\ref{obssect}.}  
\label{allspec} 
\end{figure}
Figure \ref{allspec} presents the central parts of the resulting
spectra from the four observed frequency bands in terms of main beam
brightness temperature ($T_{\rm mb}$) as a function of velocity in the
local standard of rest frame ($V_{\rm LSR}$). The continuum level for \H2O
was determined by assuming complete absorption over the 
intervals $-100$ to $-80$
and $-45$ to $+20$~\kms, giving a continuum temperature of $T_{\rm
C}$=1.29~K. This continuum level was then also adopted for \amm,
\18wat, and \wat17, since the frequencies are similar and the
continuum levels are typically not well calibrated in
position-switched data.
However, a continuum temperature determined in this way sets a definite
lower limit to the true value in our beam. It is consistent with
what we derive by interpolating between the Sgr B2 (M) dust continuum
observations at 450 and 800~$\mu$m by \citet{Goldsmith}, and similar
to the continuum temperature of 1.2~K measured by \textit{SWAS} at the
same frequency \citep{Neufeld}. 
Because the \textit{SWAS} beam was even larger ($3\farcm 3\times 4\farcm 5$)
than the Odin beam, we confirm that
the dust emission is very extended at these frequencies, filling the
Odin beam. 

The \amm\ spectrum in Fig.~\ref{allspec} shows at least seven distinct
absorption features at velocities lower than $+$60 \kms, the
approximate velocity of the Sgr B2 molecular cloud itself. The
background continuum source being at Sgr B2, all these velocity
components can be assigned to rather well-known molecular gas
concentrations along our line-of-sight towards Sgr
B2. Table~\ref{feattab} lists the absorption features and their 
origins in order of increasing Galactic radius $R$. 
\begin{table}
\newcommand\T{\rule{0pt}{2.6ex}}
\newcommand\B{\rule[-1.2ex]{0pt}{0pt}}
\caption{Absorption feature identifications in order of increasing
Galactic radius.}  
\label{feattab}
\begin{tabular}{l c l c}
\hline \hline
Feature \T & $V_{\rm LSR}$ & Origin & $R^{\rm a}$ \\
 \B & [\kms]  &  & [kpc] \\
\hline
\bf{G} \T & $+$60 & Sgr B2 molecular cloud & 0.1 \\
\bf{E}$^{\rm b}$ & $+$3 & Sgr B2 ejecta & 0.1 \\
\bf{A} & $-$107 & Expanding Molecular Ring, EMR & 0.2$^{\rm c}$ \\
\bf{B} & $-$80 & 1 kpc disk$^{\rm d}$ & $<$1 \\
\bf{C} & $-$43 & 3 kpc arm & 2.8$^{\rm e}$ \\
\bf{D} & $-$28\,/\,$-$22 & $-$23 \kms\ arm & 3.4$^{\rm f}$ \\
\bf{E}$^{\rm b}$ & $+$3 & Local gas & $\lesssim$7.6 \\
\bf{F} & $+$30 \B & $+$30 \kms\ feature & ?$^{\rm g}$ \\
\hline
\end{tabular}
$^{\rm a}$ \T Approximate galactocentric distances scaled to
$R_0=$7.6~kpc \\
$^{\rm b}$\,see discussion in text, Sect.~\ref{Res}~
$^{\rm c}$\,from H$_2$CO, \citet{Whiteoak}~
$^{\rm d}$\,see \citet{GreavesNyman}~
$^{\rm e}$\,from \ion{H}{i}, \citet{Rougoor}~ 
$^{\rm f}$\,from \ion{H}{i}, \citet{Menon}~
$^{\rm g}$\,see text Sect.~\ref{Res}.

\end{table}

To determine 
Galactocentric distances, an appropriate model of the Galactic gas
kinematics has to be applied. The most common approach among observers
has been to assume simply that the gas moves in circular orbits in the
form of a disk and/or several rings, possibly with an additional
expansion velocity to account for the 3 kpc arm feature
\citep[feature {\bf{C}}, e.g.,][]{Whiteoak}. More detailed modelling
of the large-scale Galactic gas dynamics, using a barred potential,
reproduces existing \ion{H}{i} and CO position-velocity data
\citep{Englmaier,Bissantz}, but the model resolution is insufficiently high
to make accurate predictions about the central 1 kpc of the
Galaxy. However, the model does support the idea that the
high-velocity gas ($|V|>$80~\kms) along the line-of-sight 
towards the Galactic centre is confined to this central region.

The determination of Galactocentric distance of course also depends on
the adopted Galactocentric radius of the Sun, $R_0$. Over the years,
this parameter has been assigned values between 7 and 10~kpc, and in 1985
the IAU recommended the use of $R_0=$8.5~kpc. More indirect
measurements using open star clusters \citep{Bobylev} and bulge red clump
stars \citep{Nishiyama}, as well as direct geometric measurements
of a star as close as 0\farcs1 to the Galactic central massive black
hole \citep{Eisenhauer}, all yield estimates within 0.3 kpc of
$R_0=$7.6~kpc. Thus, this value has been used to linearly scale the
distances of the references given in Table~\ref{feattab}. This also
agrees well with a measurement of the astrometric
parallax of water masers in Sgr B2, which implies that $R_0 =
7.9^{+0.8}_{-0.7}$ kpc \citep{Reid09}.

Based on both the velocity and line-shape, feature {\bf{E}} in the \amm\
spectrum corresponds to the one observed around 0~\kms\ in ammonia
inversion transitions up to (6,6) by \citet{Gardner}. These authors argue that
the high temperatures needed to excite these states, together with the
broad ($>$15~\kms) and asymmetric line-shape, imply that the gas must be
associated with Sgr B2 itself, possibly in the form of ejecta in
the direction towards the Sun. On the other hand, absorption is also
observed around 0~\kms, with line-widths of $\lesssim$10~\kms, in
\ion{H}{i} \citep{Garwood} and H$_2$CO \citep{Mehringer1995}, and
is then associated instead with cold foreground gas in the 'local'
ISM (the Orion/Sagittarius Galactic spiral arms). Furthermore,
absorption at the same velocity is also seen in the ground-state
ammonia transition in the direction of Sgr A (Sandqvist et~al., in
prep.), which would of course be highly unlikely if all of the
absorption that we observe arises in the Sgr B2 region. 

Unfortunately, the water absorptions at feature {\bf{E}}
do not contribute to an unambiguous identification of its source. Both
\18wat\ and \wat17 exhibit absorption in feature {\bf{E}} with the same
broad line-shape as that in ammonia (Fig.~\ref{allspec}), implying that the absorptions have a
common origin of a steep decline at lower velocities and then a slow
rise over about 
20~\kms. On the one hand, the high signal-to-noise ratio absorption in \18wat\
is optically thin, but nevertheless very smooth without any
overlapping narrow components that would be expected from local
gas. On the other hand, \H2O\ maps of the Sgr B2 region
\citep{Neufeld} show that the absorption around 0~\kms\ is far more
extended than that of Sgr B2 itself (corresponding to feature
{\bf{G}}), again arguing for an origin in the Galactic disk. As a consequence,
we consider feature {\bf{E}} to be a superposition of unknown proportions of
absorption from a Sgr B2 ejecta and one or several more local gas
concentrations, within a few kpc of the Sun. 

Feature {\bf{F}} has previously often been assigned to the 'Scutum
arm', about halfway between the Sun and the Galactic centre
\citep[e.g. ][]{Greaves}, but in the context of Galactic rotation models
\citep{Bissantz} this seems highly unlikely. It would mean that the
Scutum arm is falling in towards the Galactic centre, while the motion
of the well-studied 3~kpc arm is in the opposite direction. Moreover,
feature {\bf{F}} has a counterpart in H$_2$CO position-velocity
diagrams of the Galactic centre, which appears to be
localised around Sgr B2 in Galactic longitude \citep{Whiteoak}. This
is confirmed by the absence of formaldehyde absorption at the
corresponding velocity ($\sim +23$~\kms) towards Sgr~A in the study by 
\citet{Sandqvist70}, but on the other hand they  do report an
\ion{H}{i} absorption at this velocity. Because of the ambiguity of the
available data, we refrain from specifying the location of the ammonia
gas absorbing at feature {\bf{F}}.

\begin{figure}
\resizebox{\hsize}{!}{\includegraphics{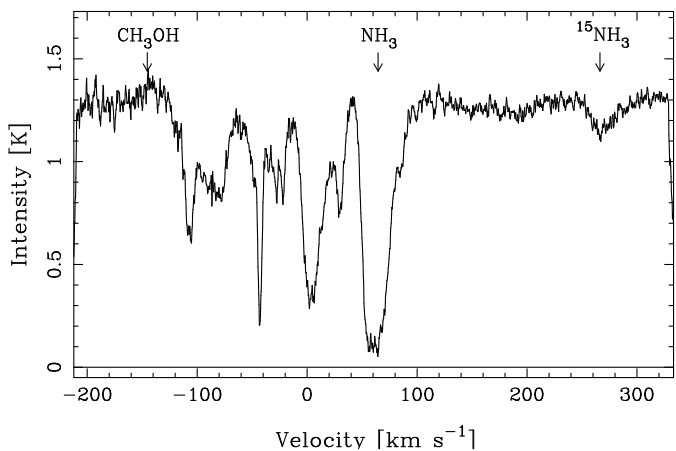}}
\caption{The full observed band covering the \amm\ absorptions towards
Sgr B2. In addition to $^{14}$\amm, the ground state $^{15}$NH$_{3}$
line, seen in absorption, and the 15$_{1,15}$\,-\,14$_{0,14}$ CH$_3$OH
transition, seen in emission, are marked at $V_{\rm LSR}$=$+$65 \kms,
respectively.} 
\label{ammspec} 
\end{figure}
The $>$1~GHz wide observed band around the \amm\ 572.498 GHz line includes
the corresponding ground-state $^{15}$NH$_{3}$
absorption at 572.112~GHz from the Sgr B2 cloud itself,
as well as emission in the CH$_3$OH $15_{1,15} \to 14_{0,14}$ transition
from that same cloud (see Fig.~\ref{ammspec}). We note that
even the most narrow \amm\ absorption features ({\bf{D}} with $\Delta
V \geq 4$~\kms) are too broad for the hyperfine quadrupole components
to be resolved ($F$\,=\,1--2, the main central component;
$F$\,=\,1--1, shifted by $-$0.6~\kms; $F$,=\,1--0, shifted by
$+$1~\kms). A straightforward line 
profile model also indicates that the slight asymmetry caused by the
hyperfine structure of absorption lines of these
intensities will be too small to distinguish -- even if the intrinsic
line-widths are as small as 1~\kms. 

The \H2O spectrum shows ground-state water absorption in the range
$-120$ to $+100$ \kms, superimposed on ground-state water emission
from the Sgr B2 cloud itself around $+60$ \kms ($\sim$20 to 120
\kms). The region around Sgr B2 was mapped in this frequency
range by $SWAS$ \citep{Neufeld}, which found that both the absorbing and
emitting water vapour is extended relative to the Odin beam. However,
the absorption lines are saturated across almost 
the whole velocity range, preventing a detailed analysis 
of the profile. 

Both \18wat\ and \wat17\ contain absorption from the Sgr B2 cloud
itself, {\bf{G}}, and feature {\bf{E}} as discussed above. In
addition, there is a clear \18wat\ absorption corresponding to the
narrow feature {\bf{C}} of ammonia (the 3~kpc arm).


\section{The line-of-sight clouds}

\subsection{Ammonia} \label{los_amm} 
As noted above, the ground-state ammonia absorption in the velocity
range $-$120~\kms\ to 40~\kms\ can all be identified with certain
molecular gas concentrations in the Galactic spiral arms and bar(s),
except the mixed feature {\bf{E}} at $V$$\approx$0~\kms\ as discussed in
Sect.~\ref{Res}. Of these, only feature {\bf{C}} has been observed in
any of the higher excited states of ammonia \citep[the (1,1) and (2,2)
inversion lines, ][]{Huttemeister93}. That absorption and no
emission is seen against a continuum background of only 1.3~K ensures
that the excitation temperature in this gas is close to $T_{\rm CMB} =
2.725$~K, the temperature of the cosmic microwave background
radiation, and that essentially all ammonia molecules reside in their
ground state. The column density of $o$-\amm\ in each component,
assuming this component to be homogeneous, can then be directly obtained
from the integrated optical depth of each feature
\begin{equation} \label{Neq}
N_{o\rm-NH_3} = \frac{\nu^3 8 \pi g_l}{c^3 A_{ul} g_u} \, \int\!\tau\,dV =
3.69 \cdot 10^{12} \, \int\!\tau\,dV, 
\end{equation}
where the column density is given in units of \percmsq. Since
$T_{\rm ex} \simeq T_{\rm CMB}$ for the (1,0)\,$\gets$\,(0,0)
transition, the optical depth ($\tau$) of each feature is also simply
given by the line-to-continuum ratio, $\tau = -\ln(1-I/I_{\rm
cont})$. In cases where we compared `features' and not simply
integrated intensities in velocity bins, we assumed that the
optical depth has a Gaussian distribution across the feature
\begin{equation}
\tau(V) = \tau_c \cdot \exp \left( -4\ln(2) \, \left(\frac{V-V_c}{\Delta V_{\rm FWHM}}\right)^2\right),
\end{equation}
where the maximum optical depth, $\tau_c$, the line velocity, $V_c$,
and the Gaussian full width at half maximum, $\Delta V_{\rm FWHM}$,
were fitted to the data in a least squares sense.

\begin{figure}
\resizebox{\hsize}{!}{\includegraphics{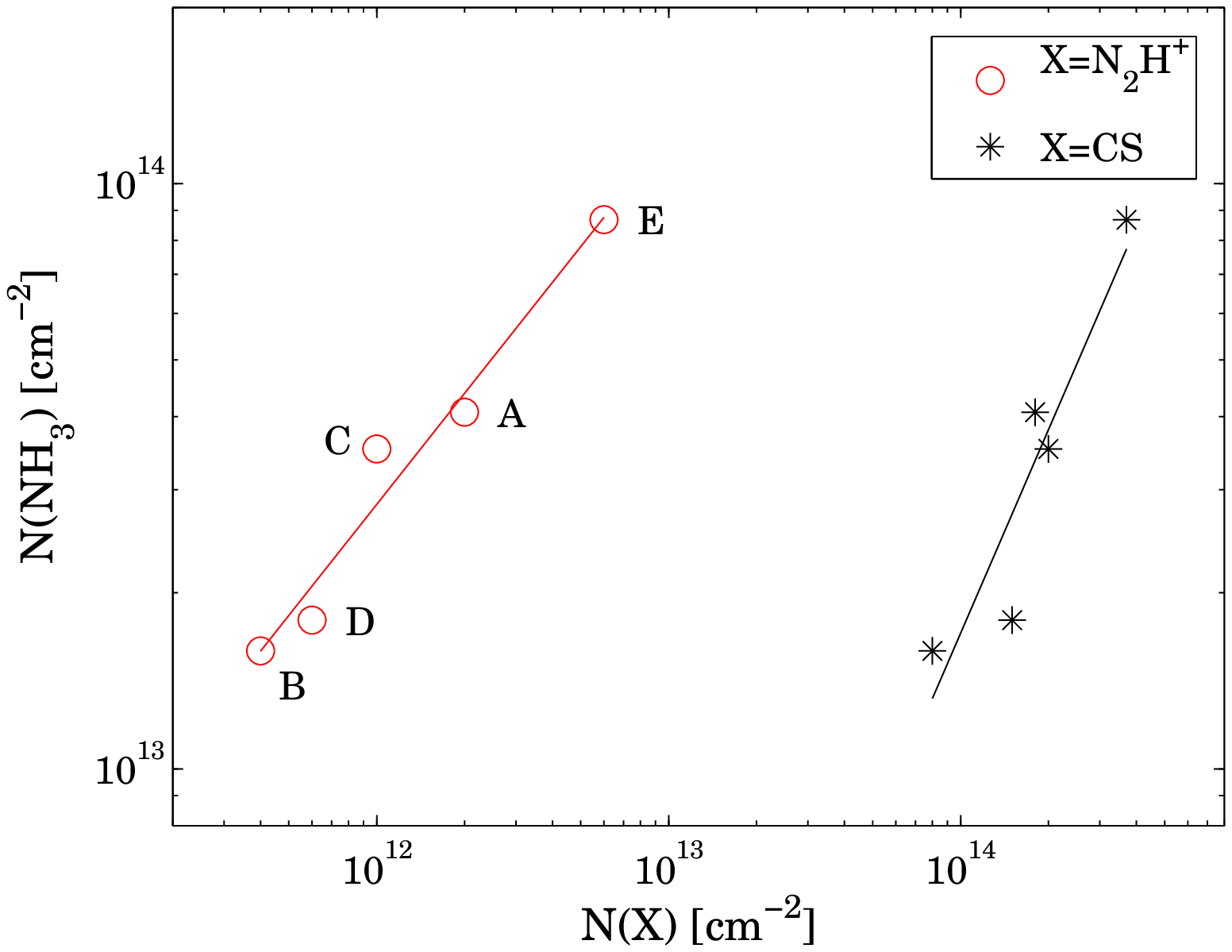}}
\caption{Column density relation between our observed $o$-\amm\ and
N$_2$H$^+$ and CS respectively. The column densities are calculated
from optical depths integrated over five velocity bins associated with
absorption features as indicated by capital letters in the
plot. N$_2$H$^+$ and CS data as well as ranges of velocity bins are
adopted from \citet{GreavesNyman}. The solid lines show the
logarithmic least-square fits to each of the data sets, and
their slopes are 1.15$\pm$0.27 and 0.63$\pm$0.07 for CS and N$_2$H$^+$,
respectively.}  
\label{CScorr} 
\end{figure}
\begin{figure}
\resizebox{\hsize}{!}{\includegraphics{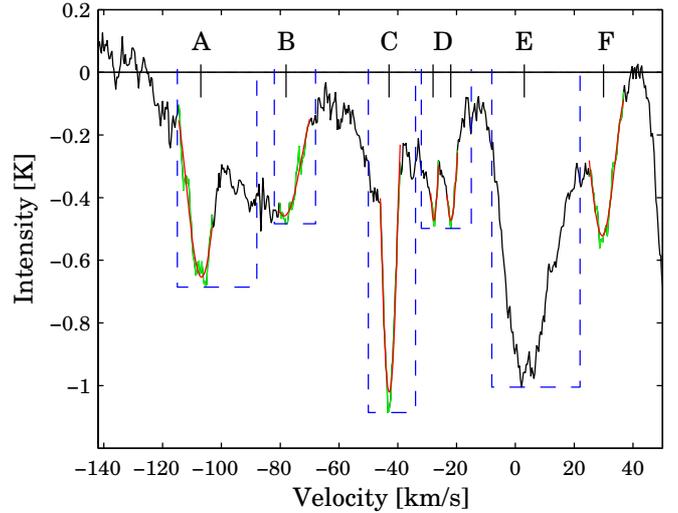}}
\caption{Gaussian optical depth fits to the observed ground-state
NH$_3$ absorptions, corresponding to components in formaldehyde. The
spectrum in black is the continuum-subtracted ($T_{\rm C}$=1.29~K)
observed spectrum. The velocity ranges over which the fits are made
are highlighted in green and the fits themselves are shown in
red. The blue dashed boxes mark the velocity ranges over which
the absorption is integrated for the column density comparisons in
Fig.~\ref{CScorr}. }  
\label{h2cogauss} 
\end{figure}

We compared the $o$-\amm\ absorption to observations of other
molecules in these 'spiral arm' clouds. For example,
\citet{GreavesNyman} observed 11 different molecules towards Sgr B2(M)
and performed a coherent analysis over six distinct velocity ranges below
$+$22 \kms. One of these velocity bins (around $-$60~\kms) was found to contain mainly
absorption in the carbon chain molecules and not to have a
counterpart in \amm. It was therefore excluded from our
analysis. The velocity bins used are indicated by blue dashed
boxes in Fig.~\ref{h2cogauss}. We also 
note that the beam-size of the \citet{GreavesNyman} observations is
smaller than half of that in the Odin observations (FWHM of 50\arcsec\  
compared to 2\arcmin\ for Odin).
Comparing our $o$-\amm\ column densities (lower limits from
Eq.~\ref{Neq}) in these velocity 'bins' to their resulting column
densities, we find striking correlations with respect to two of the 
molecules, CS and N$_2$H$^+$. Figure~\ref{CScorr} shows the
logarithmic least square fits to the data, demonstrating that there is 
a close to linear correlation between $o$-\amm\ and CS, while the
relation to N$_2$H$^+$ follows instead a square-root
law. A square-root relation such as this is what would be
expected if the increase in column density of both 
molecules is simply governed by the statistical availability of
nitrogen atoms. 
The correlation between \amm\ and CS is close to linear and implies
an abundance ratio of [CS/\amm]$\approx$3, comparable to the abundance
ratio of about 1 reported for diffuse molecular clouds outside the
Galactic plane by \citet{LisztLucas06}. The similar correlation
between CS and \amm\ implies that the chemistry in these clouds and
absorbing spiral arm clouds is not too different. 

CS was also observed towards Sgr B2(N) and (M) separately at a higher
velocity resolution (0.2~\kms) by \citet{Greaves}. Within the more
narrow velocity range covered by these observations, the large-scale
absorption structure is similar to \amm, but the number of absorption
components is higher in CS. Our interpretation is that within our much
larger beam, both encompassing (N) and (M), there are probably several
absorbing clouds at each Galactocentric distance indicated
in Table~\ref{feattab}, partly overlapping in velocities, in some
cases covering both continuum sources and in some cases not.

Early observations of ortho-formaldehyde ($o$-H$_2$CO) towards a
number of Galactic radio continuum sources, including Sgr B2
\citep{Downes,Wadiak} have approximately the same spatial and spectral
(velocity) resolution as the Odin observations \citep[beam FWHM
2\farcm1 and velocity resolution $\sim$0.2~\kms\ in][]{Wadiak}. Since
formaldehyde can be used as an H$_2$ column density tracer, with a
typical abundance of (2--4)$\times10^{-9}$ in Galactic clouds
observed in absorption \citep{Dickel90,Dickel97,LisztLucas06}, we 
use these data to estimate the $o$-\amm\ abundance in the
line-of-sight clouds. Table~\ref{abutab} presents the central optical
depths and velocity widths (FWHM) of the Gaussian optical depth fits
to $o$-\amm\ absorption components, the resulting column densities, the
corresponding centre velocities and column densities of formaldehyde,
and the resulting $o$-\amm\ abundances. We note that the fitted
components do not account for all the line-of-sight ammonia
absorption lines (see Fig.~\ref{h2cogauss}), only those that have
counterparts in formaldehyde. The reported formaldehyde absorption at
$-$45.9 \kms\ does not have a clear counterpart in ammonia. 
\begin{table*}
\newcommand\T{\rule{0pt}{2.6ex}}
\newcommand\B{\rule[-1.2ex]{0pt}{0pt}}
\caption{Results of optical depth fits and resulting $o$-\amm\
abundances derived from H$_2$CO observations by \citet{Wadiak} and CH
observations by \citet{PolehamptonCH}. Including total ammonia column
densities estimated by the representative {\tt RADEX} model of a
diffuse molecular cloud for comparison.} \label{abutab}  
\begin{tabular}{c c c c c c c c c c c}
\hline \hline
Feature \T & $\tau_{\rm NH_3}$$^{\rm a}$ & $\Delta V_{\rm NH_3}$ &
 $N$($o$-\amm) & $N$(NH$_3$)$_{\rm mod}$ & $V_{\rm LSR,\,H_2CO}$ &
 $N$(H$_2$CO) & $X$($o$-\amm)~$^{\rm b}$ & $V_{\rm LSR,\,CH}$~$^{\rm
 c}$ & $N$(CH) & $X$($o$-\amm)~$^{\rm d}$\\ 
 \B & & [\kms] & [\percmsq] & [\percmsq] & [\kms] & [\percmsq] & & [\kms] & [\percmsq] & \\
\hline
~\bf{A} \T & 0.71 & 9.8 & 2.7$\times10^{13}$ & 6.9$\times10^{13}$ & $-$105 &
$\sim$6.1$\times10^{13}$~$^{\rm e}$ & $\sim$9.0$\times10^{-10}$~$^{\rm
e}$ & $-$107.6 & 0.9$\times10^{14}$ & 1.2$\times10^{-8}$ \\
\bf{B} & 0.44 & 12.8 & 2.2$\times10^{13}$ & 5.5$\times10^{13}$ & $-$75
& $<$2.3$\times10^{13}$ & $>$1.9$\times10^{-9}$ & $-$81.7 &
2.6$\times10^{14}$ & 3.4$\times10^{-9}$ \\ 
\bf{C} & 1.57 & 4.3 & 2.7$\times10^{13}$ & 6.6$\times10^{13}$ & $-$41
& 3.3$\times10^{13}$ & 1.6$\times10^{-9}$ & $-$44.0 &
1.5$\times10^{14}$ & 7.1$\times10^{-9}$ \\  
\bf{D} I & 0.46 & 3.5 & 6.2$\times10^{12}$ & 1.6$\times10^{13}$ &
$-$27 & $<$7.0$\times10^{12}$ & $>$1.8$\times10^{-9}$ \\ 
\bf{D} II & 0.46 & 4.4 & 8.0$\times10^{12}$ & 2.0$\times10^{13}$ &
$-$20 & $<$1.2$\times10^{13}$ & $>$1.3$\times10^{-9}$ & $-$24.4 &
1.9$\times10^{14}$ & 1.7$\times10^{-9}$ \\ 
\bf{F} & 0.52 & 8.5 & 1.7$\times10^{13}$ & 4.3$\times10^{13}$ & $+$32
& 1.1$\times10^{13}$ & 3.2$\times10^{-9}$ & $+$31.4 &
1.5$\times10^{14}$ & 4.6$\times10^{-9}$ \\ 
\hline 
\end{tabular}

$^{\rm a}$ Central optical depth \T \\
$^{\rm b}$ Calculated assuming $X$(H$_2$CO)=[H$_2$CO]/[H$_2$]=2$\times
10^{-9}$ \citep{Dickel90} \\
$^{\rm c}$ Velocities from \ion{H}{i} absorption data \citep{Garwood}.\\ 
$^{\rm d}$ Calculated assuming $X$(CH)=[CH]/[H$_2$]=4$\times10^{-8}$
\citep{LisztLucasCH} \\ 
$^{\rm e}$ From \citet{Downes}, assuming $T_{\rm ex}$=3.1 K.
\end{table*}

The rather close correlation between the \amm\ and H$_2$CO column
densities, apparent from Table~\ref{abutab}, implies an $o$-\amm\
abundance of 1--3$\times10^{-9}$ in the line-of-sight clouds, if the
H$_2$CO abundance is 2$\times10^{-9}$. Assuming an ortho/para ratio
(OPR) of unity, the \amm\ abundances inferred from Table~\ref{abutab}
($X$(\amm)=3.8$\times10^{-9}$ on average) are similar to those
reported for diffuse clouds, $X$(\amm)=2$\times10^{-9}$ by 
\citet{LisztLucas06}.  

For comparison, we also used the ground-state transition of CH,
observed by ISO \citep{PolehamptonOH}, to estimate the H$_2$ column
density and the $o$-\amm\ abundance in the line-of-sight clouds. The
spectral resolution of the ISO LWS is only 30-40~\kms, but the central
velocities and velocity widths of the absorption lines were directly
adopted from the \ion{H}{i} absorption data of \citet{Garwood} and
then fitted to the data, giving estimated errors in the CH column of
typically 50\%. We adopted a CH abundance of
[CH]/[H$_2$]=4$\times 10^{-8}$, which yields overall slightly higher
$o$-\amm\ abundances (i.e., smaller H$_2$ column densities) than when 
using H$_2$CO as H$_2$ tracer (see Table~\ref{abutab}). The average
abundance for the line-of-sight clouds is then
$X$(\amm)=1.2$\times10^{-8}$, assuming an ammonia OPR of unity. 

From these two different tracers of H$_2$ column density, one mostly
giving lower abundance limits and the other having rather large
errorbars, we inferred that in general the ammonia abundance in the
diffuse clouds of the Galactic disk is of the order of 0.5--1$\times
10^{-8}$, in reasonable and unexpected agreement with the abundances
observed for diffuse clouds in the outer
Galaxy \citep{LisztLucas06}.  

The (1,1) and (2,2) inversion lines of $p$-\amm\ were observed in 
absorption towards Sgr B2 (M) at $-$40~\kms, corresponding to feature 
{\bf{C}} and associated with the 3~kpc arm, by
\citet{Huttemeister93}. A rotation temperature of 16~K was derived for
these two levels and the total column density in the two metastable
states ($\sim$1$\times10^{14}$~\percmsq, $T_{\rm ex}$$\simeq$3~K) is
significantly higher than what we find for $o$-\amm,
$\sim$3$\times10^{13}$~\percmsq\  (Table~\ref{abutab}). Even taking into
account the population in the three lowest $o$-\amm\ levels at an
excitation temperature of 16~K, the column density only reaches
$\sim$5$\times10^{13}$~\percmsq, still half of that found for
$p$-\amm. Since the velocity range of the spectra presented 
in \citet{Huttemeister93} does not cover the reported absorption at
$-$40~\kms, we were unable to estimate the possible errors in their column
density determinaton, but an OPR of less than unity is not easily
explained. The ammonia abundances that we derived for feature {\bf{C}}, using
the excitation temperature of 16~K and an OPR of unity, are
6$\times10^{-9}$ (using $N$(H$_2$)=1.7$\times10^{22}$ \percmsq\ from
H$_2$CO) or 3$\times10^{-8}$ (using $N$(H$_2$)=3.8$\times10^{21}$
\percmsq\ from CH). 

\subsection{Non-LTE excitation of ammonia} \label{radexsect}
Our measurement of the ammonia abundance was based on assumptions that have 
become standard in the astrophysical literature: that the rotational
excitation temperature relating the intensities of different inversion
transitions is closely coupled to the kinetic temperature, while the
excitation temperature relating the populations of the 
(1,0) and (0,0) levels approaches $T_{\rm CMB}=2.725$ K at low densities
in absorbing clouds \citep{WalmsleyUngerechts83,Danby88}.
The total abundance ratio of ortho to para modifications cannot be readily determined
but is presumed to ``reflect conditions at an earlier time''
\citep{Ho83}, which implies that it is related to the formation
process. An accurate accounting of the populations of many states of
NH$_3$ is needed in order to compare observations of a single
rotational line of the ortho form with one or more lines of the para
form. This accounting is greatly complicated by the existence of
metastable states, which are responsible for some of the most readily
observed absorption lines. We discuss briefly how we used detailed models of
non-LTE excitation of NH$_3$ to test the common
assumptions and refine the analysis of our observations. 

We used an enhanced version of the {\tt RADEX} code \citep{vanderTak07}
to describe the level populations and line intensities of NH$_3$ 
molecules. The molecular data included not only the rotational and inversion 
transitions, but also vibrational transitions. The energies, transition
frequencies, and transition probabilities were taken from the 
{\it ab initio} calculations of \citet{Yurchenko09}. In the current
model, 847 states and 7803 radiative transitions were considered. The
inelastic collision rates of \citet{Danby88} were adopted 
for collisions of H$_2$($J$=0) with NH$_3$ in its lowest states. In
addition, we adopted estimates of collision rates involving more
highly excited states: 
without these estimated collision rates, we would have found a large number of 
metastable and long-lived excited states with artifically high populations.
The ortho and para states were
considered together and their relative populations were controlled by
a parameter, the formation temperature $T_{\rm form}$, such that the 
chemical source rate of each rotational state $(J,K)$ was 
$$ F_{J,K} \propto g_{\rm ns} (2J+1) \exp\bigl(-E_{J,K}/kT_{\rm form}
\bigr),  $$
where $E_{J,K}$ is the energy of the state and $g_{\rm ns}$ is the 
nuclear-spin statistical weight (2 for para and 4 for ortho states).
For the chemical rates of formation and destruction expected in the 
interstellar medium, there is probably no competitive interchange process
\citep{Cheung69}. Consequently, the overall ortho/para ratio is 
controlled by the adopted value of $T_{\rm form}$. For the 
model molecule discussed here, ortho/para $\sim 1$ in the limit of 
high $T_{\rm form}$. That is, the ortho and para states contribute 
comparably to the partition function even though each ortho 
rotation-inversion level has higher statistical weight because there
are nearly twice as many low-lying para levels up to the same energy.
More importantly, in our formulation the formation process can be an
important source of highly excited metastable states, especially in 
conditions of diffuse clouds. Far-infrared continuum radiation also
plays an important role, even in diffuse molecular gas because the  
average Galactic background radiation has a significantly higher brightness
temperature than the CMB and thus excites metastable states to 
higher rotational levels more quickly than thermalizing collisions.

For a representative model of a diffuse molecular cloud at kinetic 
temperature $T_{\rm k}=30$ K and density of para-hydrogen $10^{2.5}$ 
cm$^{-3}$, we derived the following when the formation temperature is 
$T_{\rm form} = 1000$ K (formation by endoergic gas-phase processes).
The excitation temperatures of the (1,1) and (2,2) inversion transitions
were found to be 2.9 K, and those of the (3,3) inversion transition, 3.1 K, while those of 
the inversion transitions involving higher metastable states were close to
$T_{\rm CMB} = 2.725$ K. The excitation temperature of the (1,0)--(0,0)
transition at 572 GHz was found to be 2.815 K, and its optical depth
$$
\tau_{572} = 1.03 {{N/\Delta V}\over {10^{13}}}, $$
where $N$ is the total column density of NH$_3$ in cm$^{-2}$
and $\Delta V$ is the full-width at half-maximum of the line profile 
in km s$^{-1}$. The assumption of an excitation temperature close to
the temperature of the CMB is thus a good one. In the fifth column of 
Table~\ref{abutab}, the total ammonia column densities calculated from
this expression are presented for the observed line-of-sight
features. These column densities are all only slightly larger than
twice those calculated for ortho-ammonia based on the simplifying assumptions,
indicating that Eq.~\ref{Neq} and an OPR of 1 can be used as a
good approximation of a far more complicated situation.

\subsection{Water}
Since the \H2O\ absorptions in the line-of-sight clouds are highly
saturated, it is practically impossible to analyse them
directly. Only one of these clouds exhibits absorption from the less
common isotopes, that is the 3~kpc arm, feature {\bf{C}}, in \18wat. A
Gaussian optical depth fit to this feature shows that it is optically
thin ($\tau_c<$0.1), slightly broader than the corresponding \amm\ 
fit ($\Delta V_{\rm FWHM}$=6.5~\kms) and caused by an $o$-\18wat\
column density of 9.7$\times10^{11}$~\percmsq\ or more (using
Eq.~\ref{Neq} with appropriate constants). This infers a
lower limit to the $o$-\18wat\ abundance of 6$\times10^{-11}$ 
using H$_2$CO as an H$_2$ column density tracer (see end of Sect.~\ref{los_amm}). 

From the absence of absorption at this velocity in the \wat17\
spectrum, we derived a 3$\sigma$ upper limit of
$N$($o$-\wat17)$\leq$7.8$\times10^{11}$~\percmsq\ in the 3~kpc arm, 
assuming that all the $o$-\wat17\ molecules are in their ground state,
and that the line width is the same as for \18wat. Thus, adopting an
isotope ratio of [$^{18}$O/$^{17}$O]=3.5 \citep{Penzias81}, we also
derived an upper limit to the column density of $o$-\18wat in feature
{\bf{C}} of 2.7$\times10^{12}$~\percmsq. Although more extensive
surveys suggest that [$^{18}$O/$^{17}$O] varies from 2.9 near the
centre to 5.0 at $R=16.5$ kpc, the adopted value is still appropriate
for the 3 kpc arm \citep{Wouterloot}. 

The total water abundance in the 3~kpc arm is then
2$\times10^{-8}\leq X$[\H2O]$\leq$4$\times10^{-7}$, assuming 
an isotope ratio in water of [$^{16}$O/$^{18}$O]=260
\citep{Whiteoak81} and a non-LTE OPR of 1. This interval lies slightly
below the values presented for line-of-sight clouds towards Sgr B2
observed by $SWAS$ ($X$[$o$-\H2O]$\sim$6$\times10^{-7}$), but these
results are not inconsistent since feature {\bf{C}} is not clearly
distinguishable in the $SWAS$ \18wat\ spectrum due to noise and possible
baseline problems, and the $SWAS$ column has been calculated over a much
larger velocity range of 40~\kms\ \citep{Neufeld00}. Our result also
agrees with the lower limit to the water abundance in the 3~kpc arm,
$X$[\H2O]$>$2$\times10^{-9}$, obtained from the observed \H2O\
absorption towards Sgr A by Odin \citep{Sandqvist}.


\section{The Sgr B2 molecular cloud}
The absorption features around $+60$~\kms, collectively referred to as feature \textbf{G}, originating in the gas around Sgr B2 itself, are optically thick in both ammonia
and water (main isotopes). They are also the results of
superimposed absorption and emission, which is evident in the water
spectrum, but also indicated by the broad, flattish bottom of the ammonia
absorption above zero intensity (Fig.~\ref{allspec}). In addition, a
multitude of higher energy transitions of both species has been
observed at this velocity
\citep{Huettemeister95,Ceccarelli,Comito,Wilson,Cernicharo2006}, 
which is indicative of high excitation. The interpretation made by several of
these authors has been that the source geometry of Sgr B2 must include
a hot envelope or layer, with $T_{\rm kin}$=200--1000~K, outside the
warm envelope surrounding the hot cores (M) and (N). Thus, the nature
of Sgr B2 makes the analysis of feature \textbf{G} far more
complex than that of the line-of-sight absorptions.  

\subsection{\amm}
Decomposition of the \amm\ absorption feature \textbf{G} into Gaussian 
components is not unique owing to the possibly non-Gaussian intrinsic line
shapes; however, it does enable us to discern three absorbing clouds
at about +56, +65, and +85 \kms. This interpretation is in rough
agreement with maps of 6 cm formaldehyde absorption in the region,
where strong absorption at $\sim$65~\kms\ is seen towards all four
continuum peaks within the 
FWHM of the Odin beam, while the +80 \kms\ absorption only shows up
towards Sgr B2 (N) \citep{Mehringer1995}. This indicates that the
ammonia clouds absorbing at feature \textbf{G} are extended, such as those containing the water, and that the Sgr B2 (M) continuum
source is located on the near side of the +80 \kms\ cloud, while Sgr
B2 (N) is on the far side, implying a physical separation larger than
the projected one. 

Non-LTE models (see Sect.~\ref{radexsect}) for the highly excited
ammonia gas directly associated with Sgr B2 suggest an alternative
explanation of the absorption observed in highly excited metastable
states up to $(J,K) = (18,18)$. If NH$_3$ is formed by exoergic
gas-phase reactions corresponding to a formation temperature $T_{\rm
form} > 1000$ K, then the metastable levels $(J,K)$ for $J>10$ can be
populated with an apparent rotational excitation temperature $T_{\rm
rot}>600$ K without the need for a hot layer of high kinetic
temperature. Details of this scenario will be discussed elsewhere. 

\subsection{$^{15}$NH$_{3}$ and [$^{14}$N/$^{15}$N]} \label{15amm}
As mentioned above, the $^{14}$NH$_{3}$ absorption from the Sgr B2
cloud can be divided into three Gaussian optical depth
components. For the $^{15}$NH$_{3}$ absorption data, however, the
signal-to-noise ratio is not very high, and a similar
decomposition of this line is ambiguous. Nevertheless, both the
$^{14}$NH$_{3}$ to $^{15}$NH$_{3}$ optical depth ratio variation
across the absorption feature, as well as a Gaussian optical depth
fitting to the $^{15}$NH$_{3}$ line with fixed velocities and widths
adopted from the $^{14}$NH$_{3}$ components, indicate that only the
two highest velocity components are present in $^{15}$NH$_{3}$. Since
the $+$56~\kms\ component is stronger than that at $+$85~\kms\ in
the main isotopologue, this would indicate that the relative
abundances of $^{14}$NH$_{3}$ and $^{15}$NH$_{3}$ are different in
these clouds, with an unusually high [$^{14}$N/$^{15}$N] ratio in the
$+$56~\kms\ component.  

The [$^{14}$N/$^{15}$N] isotopic ratio in the solar system has been
determined to be $\sim$435 \citep{Asplund09}, and
direct measurements for nearby cold, dense clouds infer lower
ratios of $\sim$334 \citep{Lis10}. However, the ratio is not very well
known for the interstellar medium of the Galactic centre. Because of
the combination of low $^{15}$N abundance and high optical depth in
main molecular isotope lines, [$^{14}$N/$^{15}$N] has mostly been
measured in the form of a double ratio, which of course introduces
substantial uncertainites. For example, \citet{Wannier81} derived the 
ratio in the inner galaxy to be 500$\pm$100 from observations of
H$^{13}$CN and HC$^{15}$N, while \citet{Gusten} found values of
[$^{14}$N/$^{15}$N]$\sim$1000 for two molecular clouds in the Sgr A 
complex from observations of the $^{14}$NH$_{3}$ and $^{15}$NH$_{3}$
metastable inversion lines (1,1) and (2,2). \citet{Dahmen}
used additional H$^{13}$CN and HC$^{15}$N observations in the Galactic 
disk to measure a positive gradient in the ratio as a function of
Galactocentric distance. However, the previously observed Galactic
centre values were excluded from this fit, which instead inferred an
interpolated [$^{14}$N/$^{15}$N] isotopic ratio at the Galactic centre
of only 290$\pm$65. 

By fitting one Gaussian optical depth component to $^{15}$NH$_{3}$,
centred on $\sim+$67~\kms, we derived a lower limit to the column density
of $o$-$^{15}$NH$_{3}$ from Eq.~\ref{Neq} of
1.0$\times$10$^{13}$~\percmsq. For comparison, when
decomposing the absorption into two components, the lower limits would
be 8.4$\times$10$^{12}$~\percmsq\ and 1.8$\times$10$^{12}$~\percmsq\
for the $+$65 and the $+$85~\kms, respectively. The column density of
all the $^{15}$NH$_{3}$ gas giving rise to this absorption then
depends on which excitation temperature most accurately describes the population
distribution and what the OPR is. For example, if the absorption in this
transition is well described by the model for the \amm\
rotation-inversion lines observed by ISO \citep{Ceccarelli}, arising
in a hot foreground layer ($T_{\rm kin}$=700~K, 
$N$(NH$_{3}$)=3$\times$10$^{16}$~\percmsq), the [$^{14}$N/$^{15}$N]
would have to be of the order of unity, almost three orders of
magnitude smaller than expected. Thus, the major part of the 
ground-state $^{15}$NH$_{3}$ absorption is not likely to have such a
hot origin. Instead, if most of this absorption were to originate in the
warm envelope \citep[$T_{\rm kin}$=20--40~K,][]{LisGoldsmith91} with
the same column density of $^{14}$NH$_{3}$ as found by
\citet{Ceccarelli}, the [$^{14}$N/$^{15}$N] isotopic ratio would lie
between 600 and 1500 (for OPR=1), in agreement with previous
measurements. A high [$^{14}$N/$^{15}$N] isotopic ratio is expected
when the dominant nucleosynthesis process is CNO hydrogen burning,
since the slow reaction of $^{14}$N(p,$\gamma$)$^{15}$O is the one
setting the pace of the cycle.

\subsection{Water}
The water excitation in the envelope around and the core(s)
within Sgr B2 has previously been examined and modelled by several
authors. \citet{Neufeld} used the SWAS measurements of the $\rm H_2^{16}O$
and $\rm H_2^{18}O$ $1_{10}-1_{01}$ lines \citep{Neufeld00} to
show that the observed profiles could form in the envelope itself or
in a hot foreground layer. By also including HDO observations,
\citet{Comito} advocated the latter scenario. \citet{Cernicharo2006} 
extended the previous modelling efforts by using
several ISO far-IR water lines and the $3_{13}-2_{20}$ 183~GHz
line. They concluded that far-IR lines form in a hot layer but that
the 183~GHz emission clearly has to come from the colder and denser
envelope gas. Taken together, the picture is complicated and the water
ground-state lines, seen mostly in absorption, may in principle be
formed in both regions.

The water excitation and radiative transfer calculations of $\rm
H_2^{16}O$ are complicated significantly by the very large optical depths
encountered, subthermal excitation, and possibility of population
inversion. We include these non-LTE effects by using
an accelerated lambda iteration scheme \citep[ALI,][]{Rybicki91} 
to solve the radiative transfer in a spherically symmetric model
cloud. The ALI code we employed is the same as that used and
tested by \citet{Justtanont05} and \citet{Maercker08}. For the Sgr B2 cloud, we adopt the size, as well as gas and dust physical parameters for the envelope, provided by \citet{Zmuidzinas}, since those reproduce the observed continuum level around 557~GHz. The only parameter we adjusted
to help reproduce the observations was the $\rm H_2O$ abundance. We used a
turbulent velocity of 8 $\rm km\, s^{-1}$. In addition, a thin
outermost shell could have been added to mimic the hot layer. When this was
included, we adopted the parameters of \citet{Comito}.

\begin{figure}
\resizebox{\hsize}{!}{\includegraphics{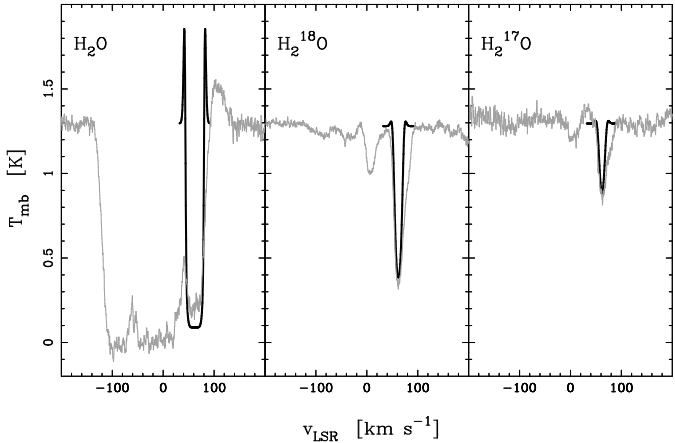}}
\caption{The ALI model results, not including the hot layer, in black,
superimposed on the Odin observations (grey) for the three ground-state
water lines. } 
\label{per_ali} 
\end{figure}

In Fig.~\ref{per_ali}, we show the model results for the ortho-water
abundances, $X[{\rm H_2^{16}O}] = 9.6\times 10^{-8}$, $X[{\rm
H_2^{18}O}] = 8\times 10^{-10}$, and $X[{\rm H_2^{17}O}] = 3.2\times
10^{-10}$. The abundances of the rarer species were determined by
matching the absorption around $+62$~\kms. The abundance of
the main species was set according to the ratio $X[{\rm
H_2^{16}O}]/X[{\rm H_2^{18}O}] = 120$, which in turn had been based upon the
$^{16}$O/$^{18}$O ratio from SO and SO$_2$ observations in the central
core of Sgr B2(M) by \citet{Nummelin}. Because 
of the high optical depth of the H$_2^{16}$O line, the abundance of the
main water isotopologue is not constrained by the model. A ratio $X[{\rm
H_2^{16}O}]/X[{\rm H_2^{18}O}] \leq 300$ cannot in principle be
excluded. However, we note that the central part of the
main water absorption profile is very sensitive to the number of
shells that the model cloud is divided into. If too few, the
absorption profile becomes too 
shallow. For a reasonable para-water abundance, $9.6\times 10^{-8}$,
this model also closely reproduces the 183~GHz observations. Hence, we
obtained the same result as \citet{Cernicharo2006} that a total water
abundance of $2\times 10^{-7}$ is adequate to explain the 183~GHz
observations. The $5_{15}-4_{22}$ 325~GHz line observed by
\citet{Melnick93} was also fitted within the uncertainties. In addition, the
same water abundance was able to fully explain the ground-state water profiles
(Fig.~\ref{per_ali}). Moreover, the ratio
$X[{\rm H_2^{18}O}]/X[{\rm H_2^{17}O}] = 2.5$ was well-constrained by
the observations, and several of the low-energy far-IR ISO lines were 
acceptably fitted by this model.

We introduced the proposed hot layer to our model by including a hot
shell ($T_{\rm k} = 500$~K, $n({\rm H_2}) = 10^3~{\rm cm^{-3}}$,
and thickness $6\times 10^{16}~{\rm cm}$) with high water abundance
outside the cloud and lowering the abundance in the envelope where the
kinetic temperature falls below 100 K. For the rarer species, we derived
good fits when adopting the ortho abundances $X[{\rm H_2^{18}O}] =
2\times 10^{-6}$ and $X[{\rm H_2^{17}O}] = 8\times 10^{-7}$ in the warm
shell. However, for a main water abundance of $(3 - 6)\times 10^{-4}$
the resulting absorption was not strong at all and mixed with
emission. As above, we found that $X[{\rm H_2^{18}O}]/X[{\rm H_2^{17}O}] =
2.5$. 

Both our ALI modelling and a direct comparison of the optical depths
of the absorptions imply an \18wat\ to \wat17\ ratio of 2.5--3 in
feature {\bf{G}}. This is lower than the mean
value for the [$^{18}$O/$^{17}$O] isotopic ratio in the Galactic disc,
3.5, as estimated by \citet{Penzias81}, but consistent with
determinations of this ratio from the three lowest rotational
states of C$^{18}$O and C$^{17}$O in Sgr B2 \citep[2.9,][]{Wouterloot}.  
This trend of a low [$^{18}$O/$^{17}$O] isotopic ratio strengthens the
argument from the high observed [$^{14}$N/$^{15}$N] (see
Sect.~\ref{15amm}) that the dominant nucleosynthesis process in the
Galactic centre is CNO hydrogen burning. The reason is that the 
reaction $^{18}$O(p,$\alpha$)$^{15}$N in the third part of the cycle
depletes the C$^{18}$O relative to C$^{17}$O
\citep[e.g.][]{Rolfs}. Furthermore, the Galactic disc is expected to
have a higher [$^{18}$O/$^{17}$O] isotopic ratio since models by
\citet{Rauscher} for massive population I stars, typical of the
Galactic disk, result in $^{17}$O production factors lower than both
that of $^{16}$O and $^{18}$O. 


\section{Conclusions}
We have presented ground-state \amm\ absorption from seven distinct velocity
features along the line-of-sight towards Sgr B2. We have identified an almost linear
correlation between the column densities of \amm\ and CS, as well as a square-root relation to
N$_2$H$^+$. Modelling the non-LTE excitation effects of ammonia in diffuse
gas, we have found that the low excitation limit, $T_{\rm ex} = T_{\rm
CMB}$, provides a good approximation of the column densities. The
ammonia abundance in the Galactic disk clouds is shown to be about
0.5--1$\times10^{-8}$. Considering the observed similarities between 
diffuse clouds in the inner spiral arms towards Sgr B2 and
clouds outside the solar circle, in terms of both ammonia abundance and the
relation between NH$_3$ and CS, a similar nature cannot be ruled out. 

From the detection of absorption in \18wat\ for the 3~kpc arm, in
combination with its absence in the \wat17\ spectrum, we conclude that
the water abundance of this arm is 2$\times$10$^{-8}$--4$\times10^{-7}$,
compared to $\sim$10$^{-8}$ for ammonia. 

The Sgr B2 molecular cloud itself is seen in absorption in \amm, \H2O,
\18wat, and \wat17, with emission superimposed on the absorption in
the main isotopologues. The non-LTE excitation of \amm\ in the
environment of Sgr~B2 could be explained without invoking an unusually
hot (200--1000~K) molecular layer in addition to the warm envelope
gas. In addition, our non-LTE ALI modelling of Sgr~B2 demonstrated that a 
hot layer is also not required to explain the observed line profiles of the 
$1_{1,0}$$\gets$$1_{0,1}$ transition from H$_2$O and its isotopologues.

We also presented ground state $^{15}$\amm\ absorption from the Sgr B2
molecular cloud. Both the high [$^{14}$N/$^{15}$N] isotopic ratio
inferred from this detection ($>$600), as well as the low \wat17\ to
\18wat\ ratio (2.5--3) in the Sgr~B2 cloud, indicate that the dominant 
nucleosynthesis process in the Galactic centre is CNO hydrogen
burning. 


\begin{acknowledgements}
Generous financial support from the Research Councils and Space
Agencies in Canada, Finland, France, and Sweden is gratefully
acknowledged. 

\end{acknowledgements} 


\bibliographystyle{aa}
\bibliography{13766}

\end{document}